# Quantifying the role of ram-pressure stripping of galaxies within galaxy groups

Tutku Kolcu[1,2]★ Jacob P. Crossett,[1,3] Callum Bellhouse[1,4] and Sean McGee[1]

[1]*School of Physics and Astronomy, University of Birmingham, Edgbaston, Birmingham B15 2TT, UK*
[2]*Astrophysics Research Institute, Liverpool John Moores University, IC2 Liverpool Science Park, 146 Brownlow Hill, Merseyside L3 5RF, UK*
[3]*Instituto de Física y Astronomía, Universidad de Valparaíso, Avda. Gran Bretaña 1111 Valparaíso, Chile*
[4]*INAF – Astronomical Observatory of Padova, vicolo dell'Osservatorio 5, I-35122 Padova, Italy*

## ABSTRACT

It is often stated that the removal of gas by ram-pressure stripping of a galaxy disc is not a common process in galaxy groups. In this study, with the aid of an observational classification of galaxies and a simple physical model, we show that this may not be true. We examined and identified 45 ram-pressure-stripped galaxy candidates from a sample of 1311 galaxy group members within 125 spectroscopically selected galaxy groups. Of these, 13 galaxies are the most secure candidates with multiple distinct features. These candidate ram-pressure-stripped galaxies have similar properties to those found in clusters – they occur at a range of stellar masses, are largely blue and star-forming, and have phase-space distributions consistent with being first infallers into their groups. The only stand-out feature of these candidates is they exist not in clusters, but in groups, with a median halo mass of $10^{13.5}$ M$_{\odot}$. Although this may seem surprising, we employ an analytic model of the expected ram-pressure stripping force in groups and find that reasonable estimates of the relevant infall speeds and intragroup medium content would result in ram-pressure-stripped galaxies at these halo masses. Finally, given the considerable uncertainty on the lifetime of the ram-pressure phase, this physical mechanism could be the dominant quenching mechanism in galaxy groups, if our ram-pressure-stripped candidates can be confirmed.

## 1 INTRODUCTION

Understanding the interrelation between galaxies and their environments is a long-standing goal of astrophysics. It has been established that galaxies in dense environments (e.g. galaxy groups and clusters) are more likely to be spheroidal and non-star-forming than galaxies in more isolated environments (Dressler 1980; Weinmann et al. 2006; Wetzel, Tinker & Conroy 2012). It is also thought that this trend is caused by galaxies being transformed or quenched as they are gravitationally drawn into these denser regions. Evidence for this comes from the redshift evolution of this environmental trend (Butcher & Oemler 1984; Balogh et al. 1999; Muzzin et al. 2012; Fossati et al. 2017; van der Burg et al. 2020), patterns in the dynamical locations of quenched galaxies (Biviano et al. 2002; Mahajan, Mamon & Raychaudhury 2011; Muzzin et al. 2014; Jaffé et al. 2015; Cleland & McGee 2021) and direct snapshots of seemingly transforming galaxies (e.g. Gavazzi et al. 2001; Kenney, van Gorkom & Vollmer 2004; Poggianti et al. 2016; Roberts & Parker 2020). Despite all this evidence, the dominant physical processes causing these transformations – and their relative dependence on environmental density (group/cluster mass), galaxy mass and redshift – are unclear.

There are at least three classes of environmentally related processes that could be causing these trends – general cosmological processes, gravitational interactions between the galaxies and their surroundings, or hydrodynamical interactions of the galaxy gas with its environment. The cosmological class of environmental mechanisms relate a broad range of potential effects such as biased formation of galaxies, the cessation of gas accretion from the surroundings or environmentally dependent outflow rates (Hearin & Watson 2013; McGee, Bower & Balogh 2014; van de Voort et al. 2017). Unfortunately, evidence for this class has come only from bulk demographics of galaxy populations and has little direct evidence as yet. In contrast, the second class of processes, gravitational interactions, have a wealth of direct evidence – mergers and galaxy–galaxy interactions as well as galaxy harassment (e.g. Barnes & Hernquist 1996; Moore et al. 1996; Moore, Lake & Katz 1998; Man et al. 2012; Deger et al. 2018). The full role of this class is not understood, partially because of the difficulty of untangling projection effects and the future evolution of galaxies in these interactions.

The third class of physical process is the ram-pressure stripping due to hydrodynamical interactions between the host galaxy and the dense intracluster/group medium (Gunn & Gott 1972). This ram pressure is a gas removal mechanism that can temporarily enhance star formation within the host galaxy by forcing the hot molecular clouds to collapse and to ignite star formation activities (Yoshida, Omukai & Hernquist 2008; Tonnesen & Bryan 2012; Kenney et al. 2014) until it effectively removes the gas and quenches the star formation completely. This mechanism can be very efficient, but depends on a combination of the galaxy's own self-gravity, the density of the surrounding gas, and the relative velocity through that gas. While there is abundant evidence that this aggressive stripping mechanism occurs in galaxy clusters (e.g. Davies & Lewis 1973; Giovanelli & Haynes 1985; Chung et al. 2009), there is relatively little evidence for it occurring in galaxy groups (Williams & Rood 1987; Rasmussen, Ponman & Mulchaey 2006; Rasmussen et al. 2008; Vulcani et al. 2018; Roberts et al. 2021b, Vulcani et al. 2021). Due to the lower densities and relative velocities, it is often thought that a gentler form of ram-pressure stripping occurs, which only removes the outer halo of gas from infalling galaxies, referred to as strangulation (Larson, Tinsley & Caldwell 1980; Balogh, Navarro & Morris 2000; Font et al. 2008; Kawata & Mulchaey 2008; McCarthy et al. 2008).

★ E-mail: t.kolcu@2020.ljmu.ac.uk

In recent years, great progress has been made in the selection and quantification of the more aggressive form of gas-stripping events. Such gas-stripping features can be identified by the 'tentacles' of gas, emission lines, and debris trails stripped from the galaxy. The long trailing features common to these galaxies has motivated the observational definition of 'jellyfish' galaxies. There is now a wide range of such galaxies identified via visual classification of imaging (both broad and narrow band) as well as from quantitative morphological measurements (e.g. Ebeling, Stephenson & Edge 2014; McPartland et al. 2016; Poggianti et al. 2016; Boselli et al. 2018; Roberts & Parker 2020). This rapid development, combined with significant IFU follow-up with the MUSE instrument and other wavelengths, has lead to a huge range of advances in our understanding of ram-pressure stripping (e.g. Fumagalli et al. 2014; Fossati et al. 2016; Bellhouse et al. 2017; Gullieuszik et al. 2017; Poggianti et al. 2017b, Vulcani et al. 2021). With some notable exceptions, the majority of this work has been focused on galaxy clusters. In this study, we aim to complement these studies by examining the properties of ram-pressure-stripped candidates (hereafter RPS candidates) in optically selected galaxy groups.

In this work, we use Hyper Suprime-Cam (HSC) *Subaru* Strategic Program imaging to identify ram-pressure stripping features of galaxies within spectroscopically selected galaxy groups from the Galaxy And Mass Assembly Survey (GAMA) within the redshift range of $0.05 \leq z \leq 0.2$. The benefits of using the HSC imaging, and the details of GAMA galaxy groups and measurements, are discussed in Section 2. A systematic selection process is established to ensure the reliability of our ram-pressure stripping classification based on the visual gas stripping evidence. The details of the selection process and the finalized catalogue are discussed in Section 3. We present our results in Section 4. We examine the stellar mass, colour, and star formation properties of the RPS candidates in Section 4.1 and examine the active galactic nuclei (AGN) content via optical line ratios (Baldwin, Phillips, and Terlevich diagram) in Section 4.2. In Section 4.3, we discuss the halo mass distribution of the galaxy groups, and in Section 4.4 we examine the phase-space distribution of the RPS candidates. In Section 5, we discuss our results with the aid of models for ram-pressure stripping and cosmological simulations, and summarize in Section 6. Throughout this work, we adopt a concordance $\Lambda$CDM cosmology, using $\Omega_m = 0.3$, $\Omega_\Lambda = 0.7$, and $H_0 = 70 \, \text{km s}^{-1} \, \text{Mpc}^{-1}$, and use AB magnitudes for all photometric measurements, unless stated otherwise.

## 2 DATA

Our study will make use of two main data sources – the spectroscopic GAMA survey and imaging from HSC. The high-completeness and depth of the GAMA spectroscopic survey are needed for the robust identification of optically selected galaxy groups, while the high-quality HSC imaging is necessary for the recovery of the low-surface-brightness features characteristic of RPS candidates.

### 2.1 Galaxy And Mass Assembly survey

The GAMA survey is a multiwavelength spectroscopic survey undertaken by using the 2dF/AAOmega spectrograph on the Anglo-Australian Telescope in New South Wales, Australia (Sharp et al. 2006; Driver et al. 2011). The survey has total coverage of 286 deg$^2$ separated in five regions as equatorial (G09, G12, and G15) and southern (G02 and G23) regions. The data used in this work is gathered from GAMA's Data Release 3 (hereafter DR3; Baldry et al. 2018), which is the first data release of the G02 region that we use. The GAMA target selection is based on two main multiwavelength imaging surveys, SDSS for equatorial regions and CFHTLS-W1 for the G02 region (Gwyn 2012) including the aid of CFHTLenS data (Heymans et al. 2012). Both SDSS and W1 observations are done in five broad-band filters (*ugriz*). In the G02 region, galaxies were targeted for spectroscopy if they had extinction-corrected magnitudes $r < 19.8$ mag.

The galaxy groups in our sample are selected from the G02 southern region that covers 55.71 deg$^2$ of the sky and has Right Ascension and Declination ranges of $30°\!.2 < \text{RA} < 38°\!.8$ and $-10°\!.25 < \text{Dec.} < -3°\!.72$, respectively. The majority of the selected region is covered by SDSS and CFHTLS-W1, and 25 deg$^2$ is overlapped with the XMM-XXL Survey (Pierre et al. 2016). Unfortunately, due to the insufficient time allocated for the GAMA survey, the G02 the region does not have high-redshift completeness for the entire G02 region. So, we prioritized the subset overlapping with the XMM-XXL survey.

The prioritized area is covered from the Declination $-6°$ and above has mean completeness 95.5 per cent where that value drops drastically to 46.4 and 31.0 per cent for the Declination ranges $-6°\!.3 < \text{Dec.} < -6°$ and $-6°\!.3 > \text{Dec.}$, respectively (Baldry et al. 2018). Therefore, galaxy groups in our sample are gathered targeting Declination $-6°\!.3$ and above, which covers a total area of 19.5 deg$^2$. The coverage of the G02 subset consists of 21 152 targets and 20 200 of those targets have a redshift quality flag, $nQ \geq 3$ (as defined in Liske et al. 2015), indicating that the sample consists of secure redshift and spectroscopic measurements. In addition to these redshift measurements, we make use of GAMA's H$\alpha$-based star formation rates (SFRs) and spectroscopic line indices for AGN classification (Gordon et al. 2018), and optical colours, as well as their spectroscopic galaxy group measurements.

As mentioned, the multiwavelength photometry data of GAMA DR3 are gathered either from SDSS or CFHTLenS input catalogues based on the specific magnitude and star–galaxy class separation described in Baldry et al. (2018). Additionally, the G02 region targeted galaxies with dust-corrected magnitudes $r < 19.8$ mag.

#### 2.1.1 Galaxy group properties

The galaxy group sample used in this work was selected from the latest GAMA Group Catalogue released in the DR3. The catalogue is constructed based on a galaxy–galaxy linking Friends-of-Friends (FoF) algorithm that uses the separation as a measure of local density. The details of the FoF selection algorithm and standard statistics and completeness corrections of groups are introduced in Robotham et al. (2011).

To select groups in the high completeness region that has Dec. $\geq -6°.0$, we select only galaxies belonging to a group with a centre above Dec. $= -6°$, to ensure high completeness for our entire sample within the G02 region. In the region above Dec. $= -6°$, there are 20 029 galaxies, with 2540 groups of two or more spectroscopic members containing a total of 8064 galaxies. However, to be assured of the group veracity, as well as having well-measured group properties including velocity dispersion and radial extents, we restricted the sample to groups with five or more members. This yields a sample of 125 galaxy groups totalling 1311 galaxies within the redshift range of $0.05 \leq z \leq 0.20$.

Within this subset, the groups range in halo mass from $10^{13.0}$ to $10^{14.5}$ M$_\odot h^{-1}$, with a mean of $10^{13.78}$ M$_\odot h^{-1}$. While our sample does contain a few structures that can be thought of as clusters (with mass $> 10^{14}$ M$_\odot$; e.g. Giodini et al. 2013; Roberts et al. 2021a), approximately 60 per cent of the structures in this subset fall below the $> 10^{14}$ M$_\odot h^{-1}$ mass. We therefore take all of these groups from the G02 subset region for further investigation.

### 2.2 HSC *Subaru* strategic program imaging

HSC is a gigantic wide-field imaging camera (with a mosaic CCD system) mounted on the prime focus of the 8.2-m *Subaru* Telescope in Mauna Kea, Hawaii with a field of view of 1°.5 diameter (Miyazaki et al. 2018). In this study, we used the second public data release (PDR2) from the HSC Subaru Strategic Program (SSP; Aihara et al. 2018, 2019). The wide component of the survey covered the GAMA G02 survey field. The wide survey contains $\sim$ 10–20 min exposures for each of *grizy* filters. The images in the *g* band (which are most useful for blue features seen in ram-pressure-stripped galaxies) have an average seeing of 0.77 arcsec and a depth of 26.6 mag. Importantly, the PDR2 contains updates to the sky subtraction methods of the survey, which no longer leads to over subtraction of a sky background correction in the outskirt regions of large galaxies on the sky. In our examination of the imaging data for jellyfish-like features, we use the *g* band images together with the multicolour composite images (*gri*).

### 3 SELECTION

Our method for identification of ram-pressure stripping features involves visual inspection of the HSC images for each of the 1311 spectroscopic group members in the redshift range of $0.05 \leq z \leq 0.20$. While full confirmation of ram-pressure-stripped features typically requires analysis of the H I or ionized gas (e.g. Chung et al. 2009; Poggianti et al. 2017b), broad-band imaging has been used to identify probable candidate galaxies in large areas of the sky (see Roberts et al. 2022, for more discussion). Optical image bands (including *u* and *B*) have been used previously to identify many ram-pressure stripping candidates (e.g. McPartland et al. 2016; Poggianti et al. 2016; Roberts & Parker 2020; Roberts et al. 2022; Vulcani et al. 2022).

In this study, we utilize the HSC *g*-band photometry as our primary diagnostic. As there are no publicly available *u*-band data from the HSC survey, we use images in the *g* band to study blue features seen in ram-pressure-stripped galaxies. While not as sensitive to the UV light as the *u* band, the image quality and depth from HSC is able to see the faint morphological features typical of ram-pressure stripping. We verify this by comparing the HSC *g*-band image of known ram-pressure-stripped galaxies to the *B*-band photometry originally used by Poggianti et al. (2016). Fig. 1 shows the *B* and *g* bands for the jellyfish galaxy JO204 (Poggianti et al. 2016). In both images ram-pressure stripping features are clearly visible, with the right image having slightly more detail. While this example is of an extreme case of ram-pressure stripping, we are confident that the imaging from HSC will be able to find features consistent with ram-pressure stripping.

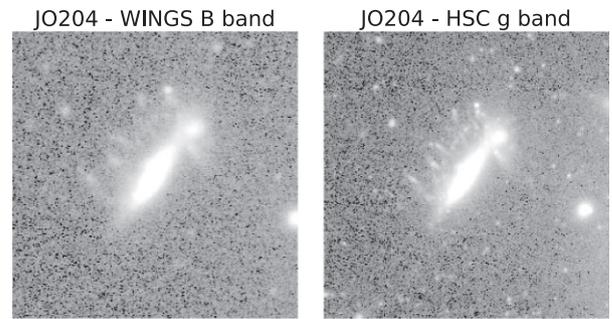

**Figure 1.** Image of jellyfish galaxy JO204, showing the WINGS survey *B*-band (left-hand panel) and the HSC *g*-band image (right-hand panel). The trails of debris due to ram-pressure stripping are seen in both images, highlighting the capability of the deep HSC imaging for identifying ram-pressure stripping features.

Each galaxy is classified by three independent human classifiers using the HSC *g*-band as well as colour-composite (*gri*) images. In this way, all galaxies in the 125 FoF galaxy groups were separately classified. Importantly, the only information (other than the images) available to the classifier was the Right Ascension, Declination, spectroscopic redshift, and the group ID. To avoid biasing the classifications, no additional information, such as which galaxy is the brightest central galaxy (BCG) or the direction of the group centre was given during the inspection.

While visual inspections are inherently subjective, we imposed a systematic list of features to search for to identify the potential ram-pressure stripping candidates. This list was intended to encapsulate any feature that may signify ram-pressure stripping in the galaxy. The selection criteria are based on the previous jellyfish galaxy studies of Poggianti et al. (2016) and Ebeling et al. (2014) and applied to each galaxy group member individually. The selection criteria specifically focused on identifying:

(i) any asymmetry and signs of deformation of the galaxy;
(ii) any evidence of debris trails;
(iii) any evidence of a tail-like structure;
(iv) a tail and/or debris trail located roughly on one side of the main galaxy body;
(v) any change of brightness around the tail appearing as knots (indicating star-forming regions);
(vi) any signs of merging activity, near-neighbours, and/or any evidence of tidal disruption

where item (vi) is included specifically to identify and exclude mergers.

We assigned a class (JF Class) to each galaxy on our list from 0 to 3 based on the strength of the gas stripping signatures defined above, particularly (i) through (v). JF Class 3 is assigned to the most confident or secure RPS candidates who carry the majority of the selection criteria. JF Class 2 galaxies are moderately confident RPS candidates that show some of the gas stripping signatures. JF Class 1 candidates carry very few of the selection criteria and cannot be confidently classified as a ram-pressure stripping candidate. JF Class 0 is assigned to galaxies for which none of the selection criteria are observed.

The presence of tidal forces does not necessarily prevent the possibility of gas stripping. Both ram-pressure stripping and tidal forces can occur at the same time as has been observed in NGC 4654 (Vollmer 2003). However, the visual evidence from HSC images was often insufficient to isolate cases where ram-pressure stripping was potentially blended with merger or tidal features. Thus, galaxies carrying any signs of merging activity or tidal interactions (satisfying criterion vi) were assigned a JF Class of −1, and were excluded from the sample and further analysis deliberately.

After excluding the mergers and ordinary galaxies, the filtered list of RPS candidates was examined by one more classifier. All contributors then discussed the results together for homogeneity and averaged the JF Class values given by three classifiers. In the remaining analysis, we define tentative and secure RPS candidates by considering their averaged JF Class. Galaxies with a JF Class average ($\overline{JF}$) ≥1.5 are classified as 'secure' RPS candidates. Galaxies with $1.5 > \overline{JF} > 0.5$ are classified as 'tentative' RPS candidates. Figs 2–4 show several example images of the different JF classes from HSC both in the *g* band and three colour. The JF Class values are specified in each image to highlight the strength of optical evidence for each assigned class.

In order to verify that our classifications are consistent with past studies, we additionally blindly classify HSC images of several ram-pressure-stripped galaxies from clusters in the WIde-field Nearby Galaxy-cluster Survey (WINGS; Fasano et al. 2006). Seven ram-pressure-stripped galaxies from Poggianti et al. (2016) were randomly included in our sample for classification. Of the seven, we assign a JF Class > 0 to five galaxies. All of these five recovered galaxies are considered to be strong cases of ram-pressure stripping (rated at JFclass ≥3 on a five-point scale in Poggianti et al. 2016), and our classification would consider these galaxies as 'secure' candidates.

We had two ram-pressure stripping galaxies that were not recovered in our analysis. Of these, one candidate had very weak stripping features, and was considered only a Jclass = 1 in P16. In our classification of this galaxy, only one classifier saw features consistent with RPS, thus not fulfilling the requirements set out in our classification criteria.

For the second galaxy that was missed by our classification, some of the classifiers deemed the asymmetries seen in this galaxy to be due to tidal interactions/mergers. This highlights that while our classifications identify several other examples of ram-pressure stripping, the subjective nature of visual classification is prone to some uncertainty.

In this paper, the strength of ram-pressure stripping was deduced only from optical images. Thus, the final classification can be dependent on the signal-to-noise ratio of the examined images, the spectroscopic redshift, and the line-of-sight galaxy orientation. The classifiers agreed in 64 per cent of the cases. However, the results from different classifiers were largely consistent in the relative ordering of the candidates with the chief difference being the overall strength of the features. It is important to note that although our list of features has similarity to other studies in the literature, the candidates are not necessarily universal. External features, such as the depth of imaging, clearly have an effect on the classifications, but also individual subjective judgement may result in different classifications. We therefore caution that the galaxies we find in this study are ram-pressure stripping candidates, and that further analysis should be conducted to fully verify the nature of these galaxies.

## 4 RESULTS

After examining the 1311 group galaxies visually, we found 45 having $\overline{JF} \geq 0.5$. This means that at least one classifier recorded the galaxy as having features consistent with ram-pressure stripping, and it was not classified as a merger/tidal remnant. For what follows, we examine the properties of the RPS candidates by dividing them into the tentative RPS candidates ($1.5 > \overline{JF} > 0.5$) and secure candidates ($\overline{JF} \geq 1.5$). We have 13 secure candidates and 32 tentative candidates.

In this section, we will compare the stellar mass, SFRs, and AGN content of the RPS candidates to the group galaxy sample. We will also examine the host halo masses of the candidates and their location in phase space.

### 4.1 Mass and star formation properties

The parent sample of galaxies in groups have stellar-masses estimated from the empirical relation between *ugriz*-derived stellar mass-to-light ratio ($M_*/L_i$) and $(g − i)$ colour as described in Taylor et al. (2011). This relation has a 1σ accuracy of ≈0.1 dex and is defined as

$$\log M_*/[\mathrm{M}_\odot] = 1.15 + 0.70(g − i) − 0.40 M_i, \quad (1)$$

where $M_i$ is the absolute magnitude in the rest-frame *i* band, expressed in the AB system.

The parent sample has stellar masses are ranging from $10^{8.00}$ to $10^{11.79}$ M$_\odot$ with a mean value of $10^{10.25 \pm 0.017}$ M$_\odot$. The secure RPS candidates have a stellar mass range of $10^{9.51}$–$10^{10.91}$ M$_\odot$ and the distribution of the tentative candidates is slightly broader. Fig. 5 shows the normalized number density histogram of the sample and shows three classes of galaxy – the parent sample of all galaxies within the galaxy groups with five or more members (black line), the secure candidates (red dashed line), and the tentative candidates (blue dotted line). There does not seem to be an obvious correlation between the stellar mass and the JF class, and the candidates are drawn from nearly the full range of stellar masses. The candidates avoid the highest stellar masses, but are otherwise distributed in a similar manner to the parent sample. The distribution of stellar masses is similar to that seen by Poggianti et al. (2016) in their sample of visually classified jellyfish.

To investigate the SFRs of galaxies, we use H α luminosity ($L_{\mathrm{H}\alpha}$), which is known to be an excellent indicator. However, to obtain reliable SFR measurements, several corrections are needed, including aperture, stellar absorption, and obscuration corrections. We obtain SFRs through the method of Gunawardhana et al. (2011), who find the following equation to return intrinsic H α luminosity from the spectral measurements:

$$L_{\mathrm{H}\alpha} = (EW_{\mathrm{H}\alpha} + EW_c) \times 10^{-0.4(M_r - 34.1)}$$
$$\times \frac{3 \times 10^{18}}{(6564.61(1+z)^2)} \left( \frac{F_{\mathrm{H}\alpha}/F_{\mathrm{H}\beta}}{2.86} \right)^{2.36} \quad (2)$$

where $EW_{\mathrm{H}\alpha}$ and $EW_c$ are the H α equivalent width and correction value, $M_r$ and *z* are the absolute magnitude and redshift value of galaxy, and $F_{\mathrm{H}\alpha}/F_{\mathrm{H}\beta}$ is the corresponding Balmer Decrement that describes the ratio of stellar absorption-corrected H α and H β fluxes (Hopkins et al. 2003). The resulting corrected $L_{\mathrm{H}\alpha}$ values are then used to derive the SFR of galaxies by using the Davies et al. (2016) relation:

$$SFR_{\mathrm{H}\alpha}(\mathrm{M}_\odot\,\mathrm{yr}^{-1}) = \frac{L_{\mathrm{H}\alpha}(\mathrm{WHz}^{-1})}{1.27 \times 10^{34}} \times 1.53. \quad (3)$$

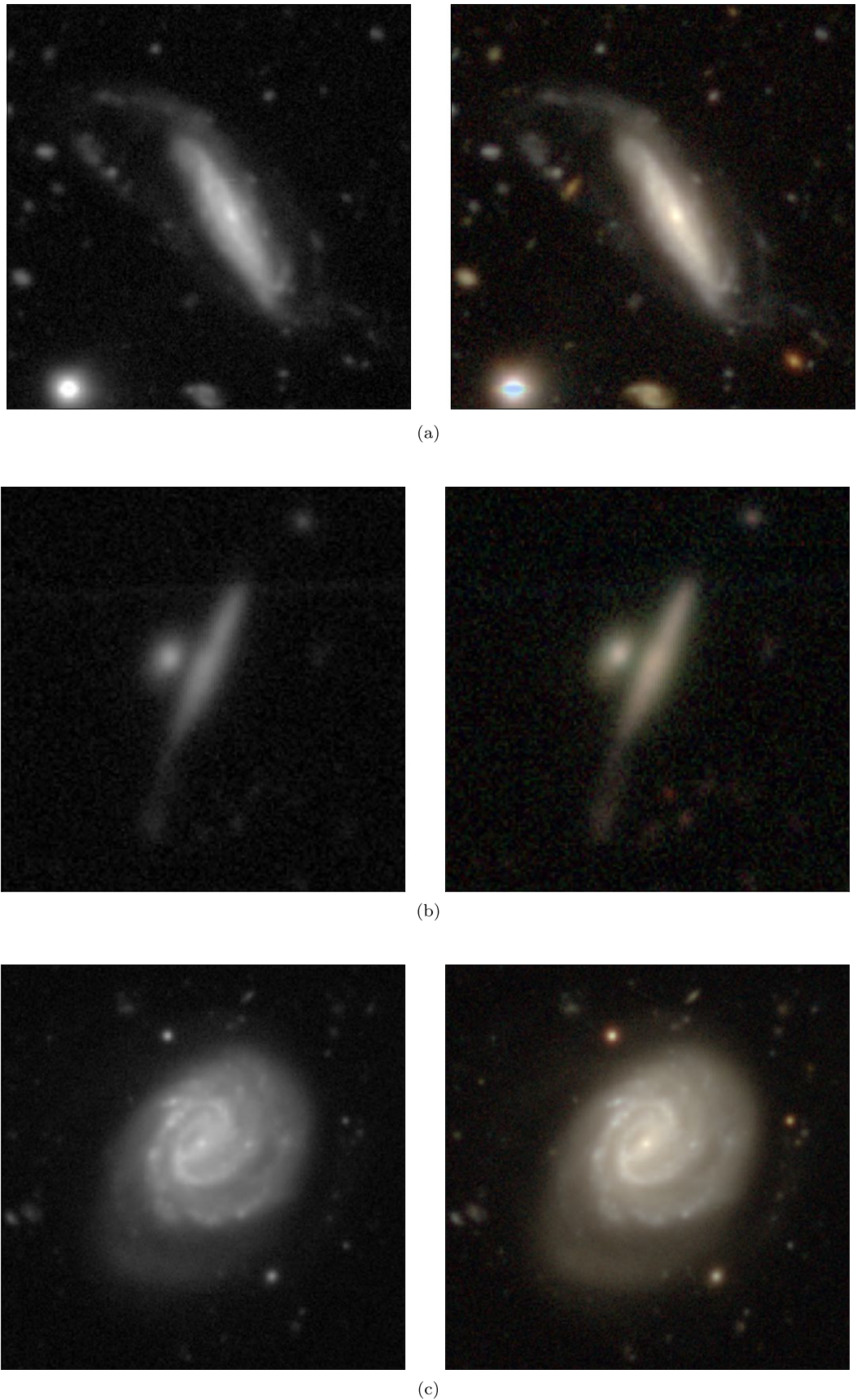

**Figure 2.** Secure RPS candidates in *g*-band (left-hand panel) and colour-composite (*gri*) images (right-hand panel). Panel (a): $\overline{JF} = 3.0$, $z = 0.08$, panel (b) $\overline{JF} = 2.3$, $z = 0.14$, panel (c): $\overline{JF} = 2.0$, $z = 0.05$.

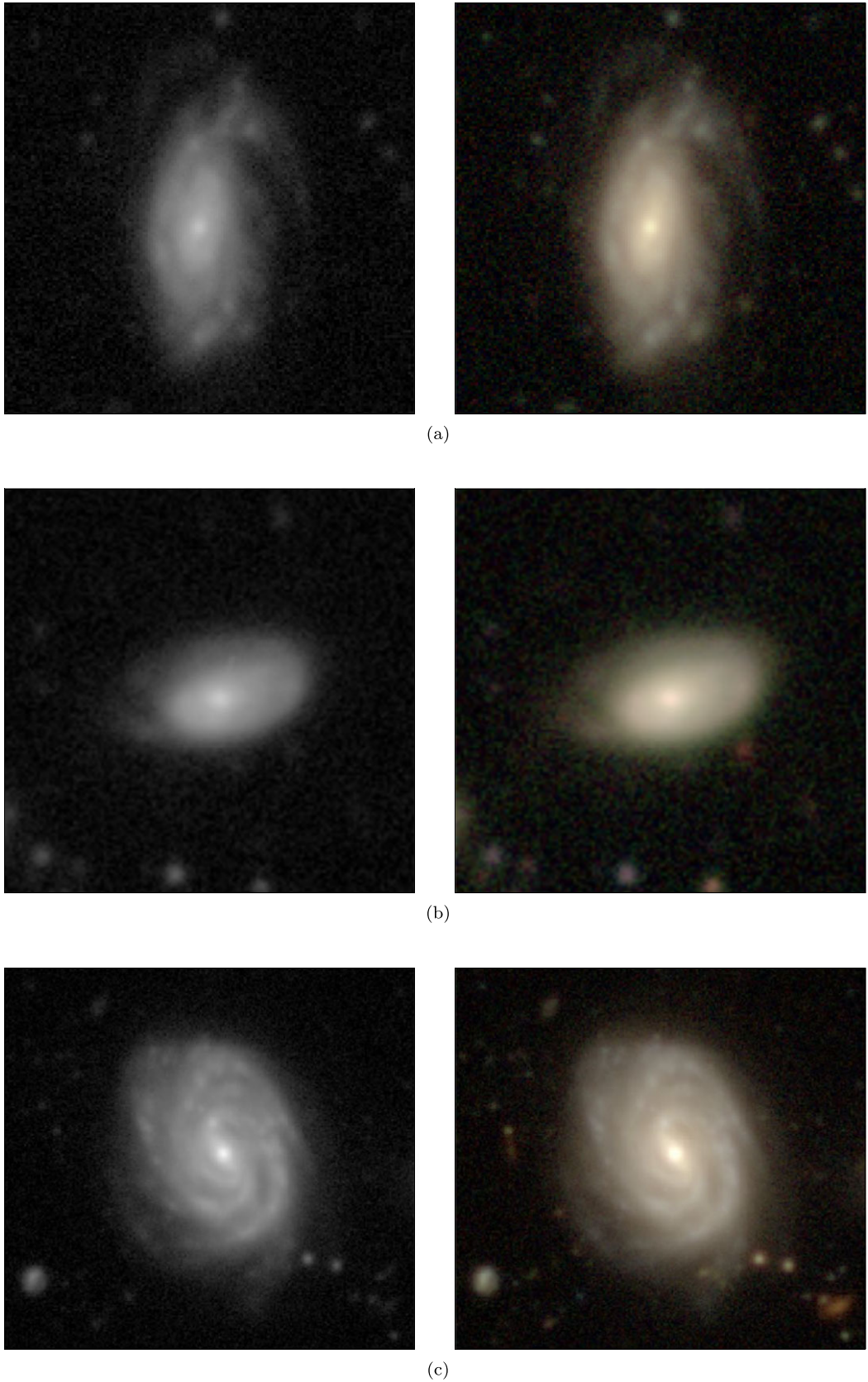

**Figure 3.** Tentative RPS candidates in *g*-band (left-hand panel) and colour-composite (*gri*) images (right-hand panel). Panel (a): JF Class average = 1.6, $z$ = 0.14, (b) JF Class average = 1.3, $z$ = 0.14, panel (c): JF Class average = 1.0, $z$ = 0.08.

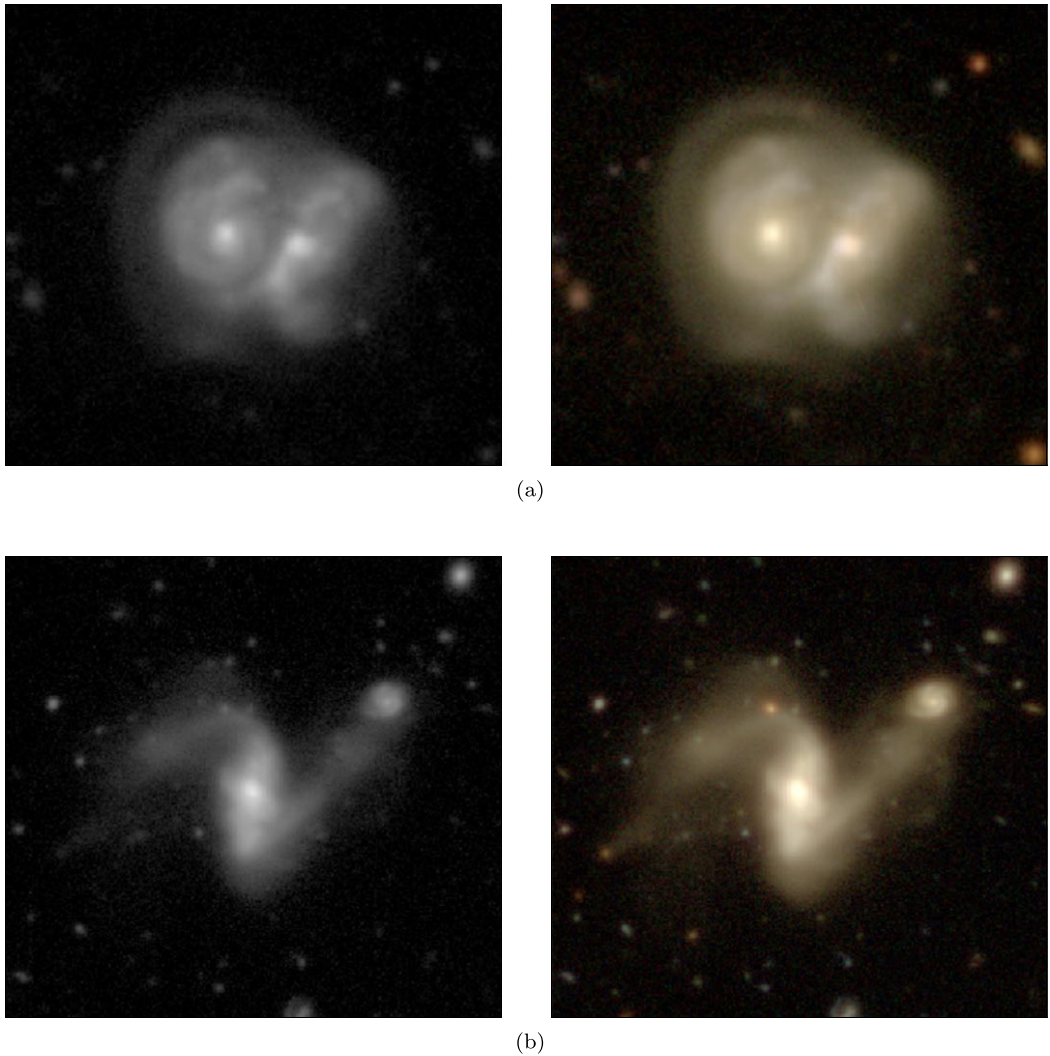

**Figure 4.** Group members classified as mergers in *g* band (left-hand panel) and three colour (*gri*) images (right-hand panel). Galaxies classified as mergers were removed from the sample. Panel (a) $\overline{JF} = -1.0, z = 0.13$, panel (b) $\overline{JF} = -1.0, z = 0.13$.

In Fig. 6, we show the specific star formation rate (sSFR) as a function of the galaxy stellar mass. The sSFR is a measure of the current to past SFRs of a galaxy. In the figure, we show the entire sample of galaxy group members, highlighting the tentative and secure RPS candidates. We separate the star-forming and passive populations using a line at sSFR $\approx -10.5$ yr$^{-1}$ similar to Davies et al. (2019).

Examining our secure and tentative RPS candidates reveals that the majority are found in the star-forming region, namely 11 out of 13 (85 per cent) of the secure candidates and 22 out of 30 (73 per cent) of the tentative candidates. This overrepresentation of RPS candidates within the star-forming population is not surprising, as these galaxies likely have significant gas that is potentially being stripped. That is, galaxies that lack gas for star formation also lack gas required for signs of ongoing gas stripping. This has also been seen in previous studies, for instance, Poggianti et al. (2016) found that most of their candidates were star-forming and in the 'blue' portion of the colour–stellar mass diagram. We compare the sSFR of each RPS candidate to the mean value for star-forming galaxies with $\Delta M_{\text{stellar}} < \pm 0.01$ M$_\odot$ of each RPS candidate. We find a 9.94 ± 0.15 and 10.08 ± 0.15 per cent SFR enhancement of secure and tentative RPS candidates, respectively, compared to the mean stellar mass-matched non-ram-pressure stripping galaxies. This shows that the SFR of our RPS candidates is enhanced compared to other group galaxies in our sample.

If we restrict the sample to only the secure RPS candidates, we notice that the only candidates with low sSFR are at the most massive end of the stellar mass distribution. This behaviour would be expected if, in order to be a star-forming galaxy, there must be a minimum gas *fraction*, while to appear as a ram-pressure-stripped candidate in a surface-brightness-limited image, there must be a minimum gas *content*. A massive galaxy may not have sufficient gas to form stars at its past average rate, but still has enough to appear as a jellyfish candidate.

Although SFRs derived from H $\alpha$ luminosities are extremely useful when accurate, they rely on corrections as well as the assumption that the origin of the emission comes from star formation rather than another source such as shocks or AGN emission. In a later section, we will examine the possibility of AGN emission, but these uncertainties motivate an examination of the colour distribution of the galaxies. In Fig. 7, we show the $g - i$ colour as a function of mean stellar mass. This broad-band colour has the advantage of simplicity and lack of aperture corrections, but the drawback that dust obscuration can make a galaxy appear redder than expected from its otherwise intrinsic properties. To create a demarcation between blue (star-forming) and red (passive) galaxies, we fit the red-sequence line by performing a non-linear least-squares fit for galaxies with $g - i > 0.95$, then move the line by 0.2 in the $y$-axis (colour) in $g - i/M_{\rm stellar}$ plane. We find a separation line between the red sequence and blue cloud to be

$$g - i = 0.090 \times M_* + 0.020, \quad (4)$$

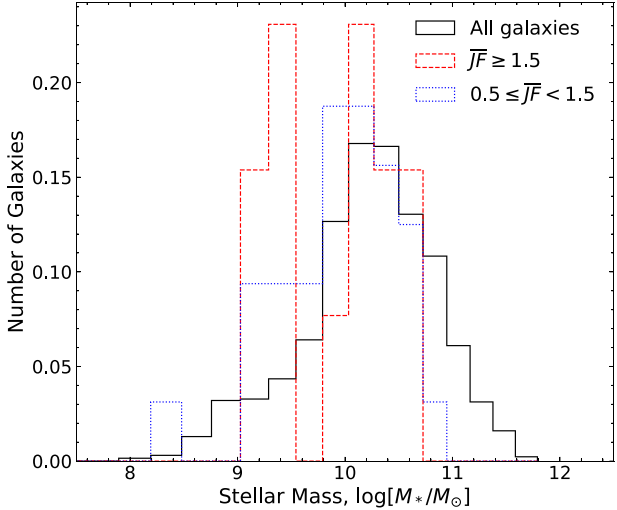

**Figure 5.** Distribution of galaxy stellar mass for three subsets of galaxies. Each set is normalized so that the area equals one. All galaxies within the galaxy groups with five or more members are shown in the solid black line, the secure ram-pressure candidates $\overline{JF} \geq 1.5$) have the red dashed line, and tentative candidates ($0.5 \leq \overline{JF} < 1.5$) have the blue dashed line.

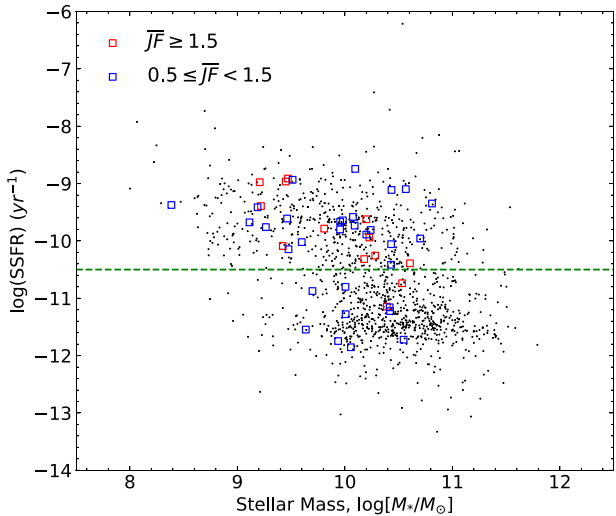

**Figure 6.** sSFR as a function of galaxy stellar mass. The dashed green line represents the separation limit taken as sSFR $= -10.5$. All galaxies in groups of 5 or more members are shown as black dots, while secure (red squares, $\overline{JF} \geq 1.5$) and tentative (blue squares, $0 < \overline{JF} < 1.5$) RPS candidates highlighted separately.

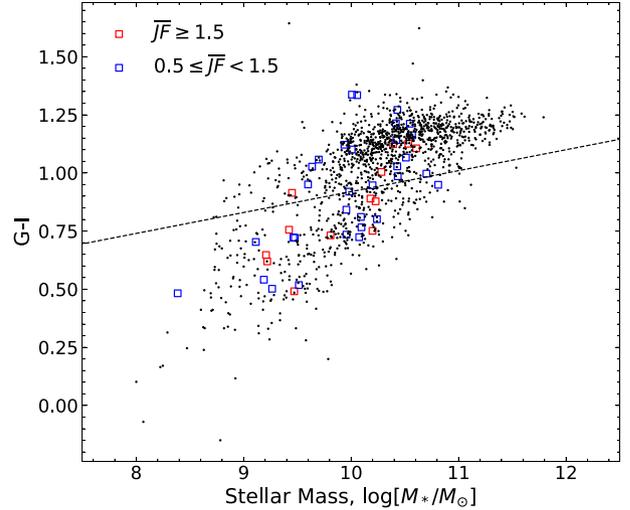

**Figure 7.** $g - i$ colour–stellar mass distribution of all galaxies within groups of five or more members in the sample. All galaxies are shown as black dots, while secure (red squares) and tentative (blue squares) RPS candidates are highlighted separately. The dashed black line is a demarcation between blue and red galaxies that was found by offsetting the fit red-sequence line.

where $M_*$ corresponds to the stellar mass of each galaxy.

Mimicking our results using the H $\alpha$ luminosities, the majority of the RPS candidates are located in the blue cloud (and thus star-forming) region of the colour–mass diagram. It is worth reiterating that our visual selection did not take the star formation/colour properties directly into account, and thus, their RPS candidates appearance in a preferential location in the colour and star formation planes is indicative both of their true nature and evidence of the fidelity of our classifications. Although, it is worth noting that the occurrence of blue stripping tails could make an otherwise red galaxy appear somewhat bluer.

### 4.2 AGN features

AGN are the compact central region of galaxies that are theorized to be powered by material accretion on to the core supermassive black hole. It is thought that some physical mechanisms such as tidal disruptions, merging, harassment, and ram-pressure stripping can create an imbalance in the interstellar gas reservoir and introduce more cold gas to feed the AGN (Sanders et al. 1988; Moore et al. 1996; Gatti et al. 2016; Poggianti et al. 2017a, Peluso et al. 2022). In this section, motivated by the findings of Poggianti et al. (2017a), who found that six out of their sample of seven jellyfish galaxies had strong AGN emission, we search for signatures of AGN emission in our sample of ram-pressure stripping candidates.

A popular method for ascertaining the AGN content of galaxies from optical emission lines uses the Baldwin, Phillips & Terlevich (BPT) diagram (Baldwin, Phillips & Terlevich 1981) that we present in Fig. 8, and again we have plotted the full sample of group galaxies as well as the tentative and secure RPS candidates.

To quantify the number of AGN-dominated RPS candidates in our catalogue, we use the Kewley et al. (2001) line. Kewley et al. (2001) used SPS and photoionization models of starburst galaxies to determine the maximum photoionization capable from star formation and their locus on the BPT diagram (shown as a dashed green line in Fig. 8). As this line is the maximum that can occur from star formation, galaxies above this line are dominated by AGN emission. Interestingly, only two of our RPS candidates are classified as AGN dominated by this demarcation, and both of those are tentative candidates. So, 0 out of our 13 secure candidates and only two out of our 30 tentative candidates are Kewley-classified AGN.

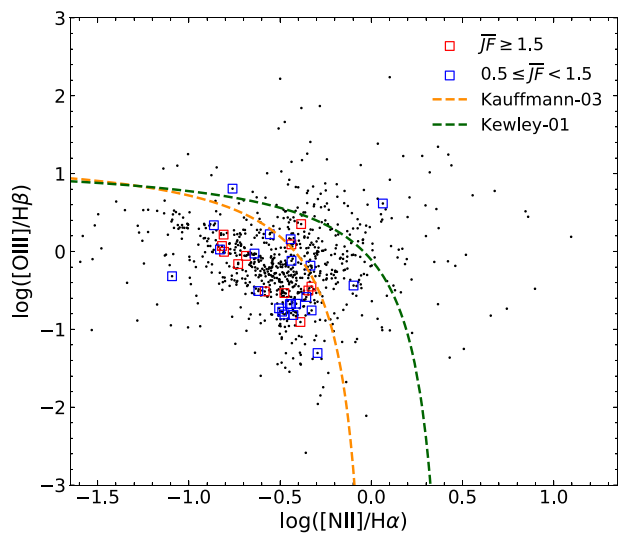

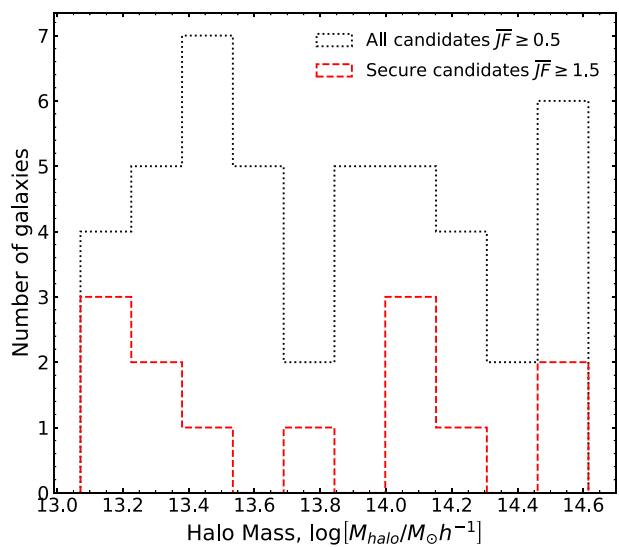

**Figure 8.** BPT diagram. All galaxies in groups of five or more members are shown as black points, while secure and tentative RPS candidates are highlighted as red and blue squares, respectively. The dashed orange and green lines represent the star-forming–AGN separation lines of Kauffmann et al. (2003) and Kewley et al. (2001), respectively. The area between both lines defines the composite region. The region above the Kewley-01 line is defined as AGN-dominated region and the region below the Kauffmann-03 line is the pure star-forming region.

The separation line of Kewley et al. (2001) is a strict demarcation, and there are known 'composite' galaxies like Seyfert-H IIs that contain both star formation and AGN features in their spectra. To distinguish such objects from pure star-forming galaxies, we used the modified Kewley line from Kauffmann et al. (2003) (which we call Kauffmann-03 line, shown as the orange dashed line in Fig. 8).

Similar to the results using the classification of Kewley et al. (2001), the majority of our RPS candidates (both secure and tentative) are below the AGN classification of Kauffmann et al. (2003), which indicates a pure star-forming galaxy. As expected, the star-forming region shows a tight clustering from low to high metallicities on the BPT diagram. Inclusive of the Kewley AGN-dominated candidates, we find that a total of 7 out of the 43 candidates are above the Kauffmann demarcation (2 of the 13 secure candidates and 5 of the 30 tentative candidates).

### 4.3 Halo mass distribution of RPS candidates

Compared with clusters, galaxy groups have a less dense intragroup medium, and the member galaxies within groups have lower relative velocities. This makes the conditions in groups suboptimal for efficient gas stripping compared to clusters, as the ram pressure depends on both the density of the ICM and the relative velocity between cluster and galaxy (Gunn & Gott 1972). Therefore, it is intriguing that our ram-pressure-stripped candidates are found across a large range of halo masses.

We can examine the host halo masses using halo mass estimates calculated in Robotham et al. (2011). These estimates are a measure of the dynamical mass and are calculated from the velocity dispersion and spatial extent of the galaxy group, with a calibration factor determined by comparison to numerical simulations. In Fig. 9, we show the histogram of halo masses of our ram-pressure-stripped candidates (both the tentative and secure). Both the tentative and secure candidates are found across a surprisingly wide range of halo masses, from $10^{13}$ to $10^{14.6}$ $M_\odot$.

**Figure 9.** The halo mass of groups hosting RPS candidates. The black dotted line represents the all RPS candidates ($\overline{JF} \geq 0.5$), where the red dashed line represents the secure candidates only ($\overline{JF} \geq 1.5$).

The traditional halo mass boundary between a group and a cluster is $10^{14}$ $M_\odot$ (e.g. Giodini et al. 2013). We find that there are > 60 per cent (28 of 45) of our candidates in systems with halo masses usually called groups with the remaining hosted by clusters. Similarly, 7 of the 13 (53 per cent) secure candidates are also hosted in group sized haloes. This shows that groups as well as clusters are capable of inducing ram-pressure stripping features in galaxies. We will show in section Section 5.3 that ram pressure is expected in at least some galaxy groups.

### 4.4 RPS candidates in phase space

The phase-space distribution of galaxies within groups or clusters has shown to be an effective indicator of the accretion state of the population – both in simulations (Oman, Hudson & Behroozi 2013; Muzzin et al. 2014; Rhee et al. 2017) and observations (Mahajan et al. 2011; Noble et al. 2013; Muzzin et al. 2014; Jaffé et al. 2015). In this context, phase space refers to the position in group-centric radius (e.g. how far is the galaxy from the centre of the group) and the relative velocity (e.g. what is the redshift-determined velocity offset between the galaxy and the group centre). Galaxies with preferentially high group-centric radius or velocity offset are likely to have been more recently accreted into the group than galaxies with lower velocities and radii. Thus, this measure can separate galaxy classes such as infalling (galaxies on their first passage into the group), back-splashing (galaxies that have already orbited the group centre, and may have left the group briefly) and the stable, virialized galaxy populations that have been in the group for several Gyr.

In this section, we use the observed group velocity dispersion values as calculated with the methods of Robotham et al. (2011). To calculate $R_{200}$, we combine the velocity dispersion–mass relation as determined by Munari et al. (2013),

$$\sigma_{1D} \approx 1100 \text{ km s}^{-1} \left[ \frac{h(z) M_{200}}{10^{15} \text{ M}_\odot} \right]^{1/3}, \quad (5)$$

with the relation between $M_{200}$ and $R_{200}$ given by Durret et al. (2021),

$$GM_{200} = 100 H(z)^2 R_{200}^3. \quad (6)$$

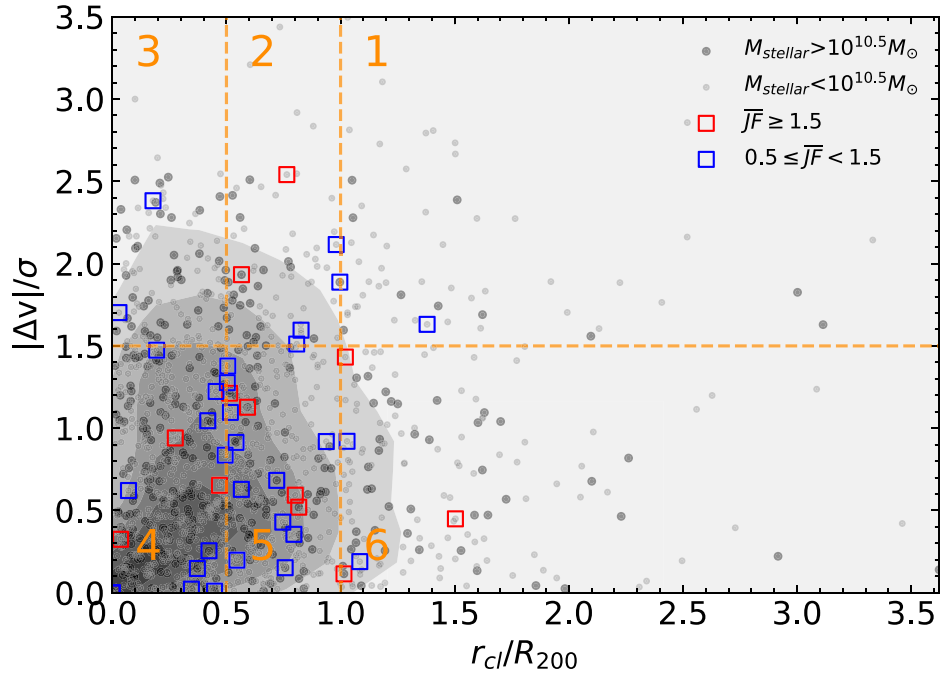

**Figure 10.** The phase velocity diagram of stacked galaxy groups within the sample. Galaxies with stellar mass higher and lower than $10^{10.5}$ are marked as black and grey points, respectively. The contour lines represent the number density of galaxies combined (darker colour indicates a higher population). The secure and tentative RPS candidates were highlighted as empty red and blue square markers respectively. The diagram is divided into six regions by orange dashed lines, following a similar procedure to Roberts et al. (2021a,b), aiming to highlight the RPS candidate excess in each region (Fig. 11).

We calculate the line-of-sight velocity from the relation between redshift and velocity dispersion ($\sigma$) as

$$\frac{|\Delta v|}{\sigma} = \frac{|c(z_{gal} - z_{group})|}{(1 + z_{group})\sigma}, \quad (7)$$

where $z_{gal}$ and $z_{group}$ are the redshifts of the individual galaxy and the group centre respectively. The $(1 + z_{group})$ factor is needed to correct the velocity to the rest frame.

We show the resultant phase-space diagram in Fig. 10 for both the secure and tentative RPS candidates. We also show the location of all the group galaxies that were examined to show that the RPS candidates are not simply drawn randomly from the set of all galaxies. The strong RPS candidates seem to be located at preferentially high relative velocities and relatively low $R_{200}$ values, consistent with the expected locations for galaxies experiencing ram pressure.

We confirm this by calculating the excess number of ram-pressure-stripped candidates in different regions of the phase-space diagram. We divide the phase-space plot into six regions, shown in Orange on Fig. 10. We compare fraction of ram-pressure-stripped galaxies in any given region, to the fraction of the total galaxy population within that region following the procedure used in Roberts et al. (2021a,b).

The formal definition is given as

$$\text{RPS candidate excess} = \left( \frac{N_{RPS}^{R_i}}{N_{RPS}} \right) \bigg/ \left( \frac{N_{Group}^{R_i}}{N_{Group}} \right), \quad (8)$$

where $N_{RPS}^{R_i}$ is the number of RPS candidates in region $R_i$, $N_{RPS}$ is the total number of RPS candidates (both secure and tentative), $N_{Group}^{R_i}$ is the number of GAMA group members in the region $R_i$, and $N_{Group}$ is the total number of GAMA galaxies. While the method here follows that of Roberts et al. (2021a,b), our analysis uses the group $R_{200}$ instead of $R_{180}$. Additionally, our group membership extends beyond $1 \times R_{200}$. We therefore add two extra regions beyond $1 \times R_{200}$ to account for the extra galaxies beyond this radius. The regions are separated at $0.5 \times R_{200}$ and $1 \times R_{200}$, as well as $1.5 \times (|\Delta v|/\sigma)$.

We show the ram-pressure stripping excess for each region in Fig. 11. We see that there is a $\gtrsim 1\sigma$ excess in the phase-space region 2, corresponding to high relative velocities galaxies within $R_{200}$. There is also a $\gtrsim 1\sigma$ excess in region 5, where galaxies may still be infalling (albeit with lower line-of-sight velocities). This further explored in Section 5.2. Interestingly, the secure RPS candidates also tend to avoid the central regions of the phase-space diagram (region 4 – with low velocity and low $R$). This is similar to the avoidance of the central regions seen in post-starburst galaxies in galaxy clusters (e.g. Muzzin et al. 2014). We note that these results have large uncertainties, as seen in Fig. 11, due to the modest sample size in each bin.

We will return to the discussion of phase space in Section 5.2, but it is worth pointing out that our results are consistent with several other studies. For instance, Jaffé et al. (2018) found that jellyfish galaxies in their sample had higher relative velocities and lower cluster-centric radii than the typical galaxy in their clusters. The result is also similar to the results for radio-selected RPS galaxies in galaxy clusters as in Roberts et al. (2021b). However, we see differences between the our result and the flat distribution of RPS excess found within galaxy groups in Roberts et al. (2021a). It seems that our ram-pressure-stripped candidates in groups are consistent with the properties of jellyfish galaxies seen in galaxy clusters, albeit with a low significance. Future studies, combining larger sample of ram-pressure-stripped candidates, may be able to better constrain the infall properties of ram-pressure-stripped galaxies in groups.

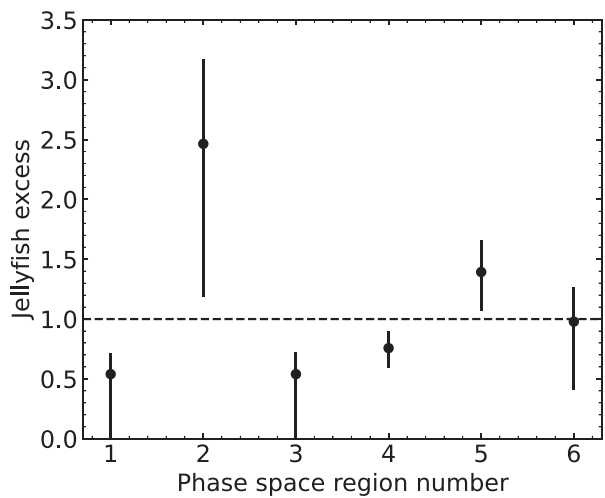

**Figure 11.** Excess fraction of ram-pressure stripping candidates compared to the number of group members in six different regions of the phase-space diagram. 1$\sigma$ binomial uncertainties are calculated following Cameron (2011). We see the highest excess of ram-pressure stripping candidates is found at high velocity offsets, but within $1 \times R_{200}$.

### 4.5 Tail orientations

One of the more striking features common to many ram-pressure-stripped galaxies are the 'tails' of star formation and gas that trail behind the galaxy. These tails are expected to follow the galaxy as they infall, and have been used in the past as probes of the direction of motion of an infalling galaxy (e.g. Chung et al. 2007; Roman-Oliveira et al. 2019; Roberts et al. 2021b). In observational studies of ram-pressure-stripped galaxy populations, tails are generally found to be pointing away from the cluster centre, suggesting that the galaxies are often on first infall (e.g. Smith et al. 2010; Roberts & Parker 2020). The lack of galaxies with tails perpendicular to the cluster core could also suggest that these galaxies are on more radial orbits, and potentially experiencing maximum ram pressure (Roberts & Parker 2020; Roberts et al. 2021b). These results also depend on the dynamical state of the host group or cluster, with cluster mergers washing out the preferential infall direction (Roman-Oliveira et al. 2019), and even causing tails to align with the cluster merger axis (Rawle et al. 2014).

While there are limitations to inferring the direction of motion of a galaxy using ram-pressure stripping debris tails (e.g. Roediger & Brüggen 2006), this method can be used to further investigate the infall properties of our sample. We attempt to assign tail directions to all 45 of our ram-pressure stripping candidates, following a procedure similar to other optical studies (e.g. McPartland et al. 2016; Roberts & Parker 2020). Three classifiers all attempted to individually estimate the direction of the tail (or in the absence of a tail, the estimated direction of motion) from three-colour image cutouts of each candidate. No other information about the galaxy, including the location of the group centre, was given during the classification. The three votes were then compared, and used to generate an average tail direction.

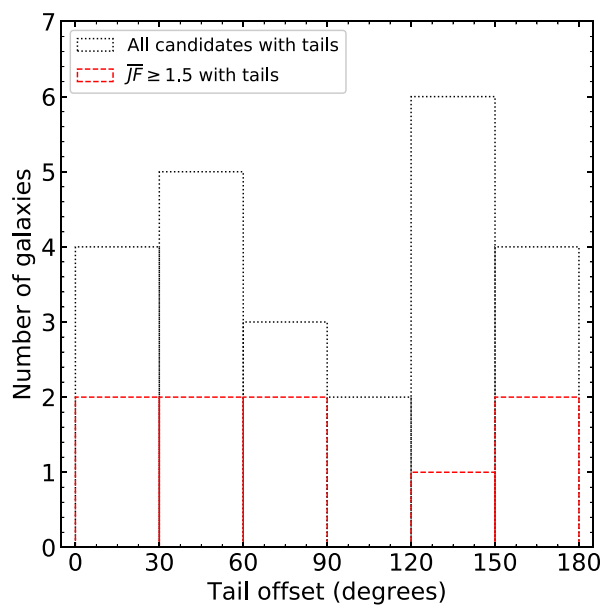

**Figure 12.** Histogram of the direction of tails of all RPS candidates (black dotted line), and only the secure candidates (red dashed line), compared to the group centre. Galaxies with a tail pointed towards the cluster have an angle of 0°, while tails pointing away from the centre are 180°. We see a deficit of galaxies with angles between 60° and 120°. This suggests that many of the galaxies may be on infalling or backsplashing group orbits, consistent with other ram-pressure stripping galaxies.

In 27/45 cases, at least two classifiers were able to estimate the tail direction within 45° of each other. Of these, all three classifiers agreed on the tail on 14 occasions, while only two classifiers were able to detect a tail in 11 cases. In two specific cases, two classifiers agreed on a tail angle, while a third differed. In these cases, the single discrepant measurement was disregarded, and a new average calculated. These average values were then compared to the direction of the BCG, to get a final tail angle. Note in the following, a tail angle of 0° denotes a tail pointing towards the group centre, and 180° points away from the centre.

Fig. 12 shows the distribution of tail angles for 24 of our RPS candidates with estimated tails, and which are not considered to be at the centre of a group. The entire sample is shown in the black dotted line, while the secure RPS candidates are shown separately with a red dashed line. We see there is a deficit of galaxies with tail angles between 60° and 120°, compared with tails pointed either towards the cluster centre (<60°), or away from the group centre (>120°).

The extra tails pointed away from the group centre provides possible evidence that these galaxies are on radial orbits, and may be infalling into the group for the first time. This result mirrors that seen in large clusters, where tails of ram-pressure-stripped galaxies are preferentially pointed away from the cluster (e.g. Chung et al. 2007; Smith et al. 2010; Roberts et al. 2021b). However, we also see an excess of RPS candidates with tails pointed towards the cluster. These galaxies have potentially passed infall, and are now backsplashing after their first passage through the cluster. This matches the distribution seen in Roberts et al. (2021a), who reported a similar spread of tail angles in ram-pressure-stripped galaxies in group environments. They argue that the lower ICM densities, as well as lower infall velocities of galaxies in groups can delay the stripping process until later in their infall, resulting in a population of galaxies with tails pointing towards the group centre in addition to those infalling. The distribution we see in Fig. 12 supports this scenario. We further explore the ability for groups to strip gas on first infall in Section 5.3.

# 5 DISCUSSION

It could be said that several of our results are not surprising. That is, the RPS candidates come from a range of stellar masses, and are predominately blue and star-forming. Our ram-pressure stripping candidates share many properties with stripped galaxies from other studies, with one caveat – these are candidates found in galaxy groups, not clusters. Indeed, the similarity of these types of ram-pressure-stripped galaxies suggests the process is very similar in groups and in clusters. The reduced velocities and less dense ICM that characterize galaxy groups are still able to produce morphological features in galaxies consistent with ram-pressure stripping. In this section, we discuss a few of the more intriguing aspects of our results – the lack of AGN content, the existence of ram-pressure stripping candidates in groups, and their locations in phase space. Finally, we discuss the existence and prevalence of these candidates in groups, and what role ram-pressure stripping might play in the environmental effects observed on galaxies in general.

## 5.1 On AGN fueling

As we saw in Section 4.2, very few of the RPS candidates and none of the secure candidates have emission-line ratios consistent with being dominated by AGN emission. We may have expected to see some AGN emission fuelled by the stripping events. In Poggianti et al. (2017a), it was found that galaxies experiencing strong ram-pressure stripping had very high incidence of the presence of AGN emission (six out of seven galaxies). Further, McGee (2013) found that satellite galaxies had larger black holes for their stellar mass, consistent with having preferential black hole accretion. Evidence for this ram-pressure-stripped-induced black hole accretion has also been found in numerical simulations (Ricarte et al. 2020), although others do not see an enhancement of black hole accretion in the general population of disc galaxies in clusters (Joshi et al. 2020). Indeed, further numerical simulations suggest that some of the overly massive black holes are caused by tidal stripping of stellar material, and not via ram-pressure-induced accretion (van Son et al. 2019).

In Poggianti et al. (2017a), the authors examined the most strongly stripped galaxies in a large sample of RPS candidates, which had confirmed H$\alpha$ tails at least as large as the stellar disc. This is potentially a large difference from our study as we may not have any such candidates in our sample, either because such extreme examples do not exist in galaxy groups or because they are simply rare in general and would require a larger sample to find some. Further follow-up of our sample is needed to determine which of these is the case – suffice to say, our results are not in conflict with those of Poggianti et al. (2017a) because of these sample differences.

## 5.2 Simulated galaxies in phase space

As we have discussed in Section 4.4, the position of galaxies in phase space can be a clear indicator of their accretion state – that is, how long it has been since they have entered the group or cluster. In Fig. 10, we found that the RPS candidates preferentially had higher velocity offsets at a given group-centric radius than the rest of the group population. Furthermore, they seemed to avoid the origin of the phase space, where most of the virialized galaxies are expected to reside. To get a better quantitative idea of the implications of this observed offset we compare to the results of a semi-analytic simulation.

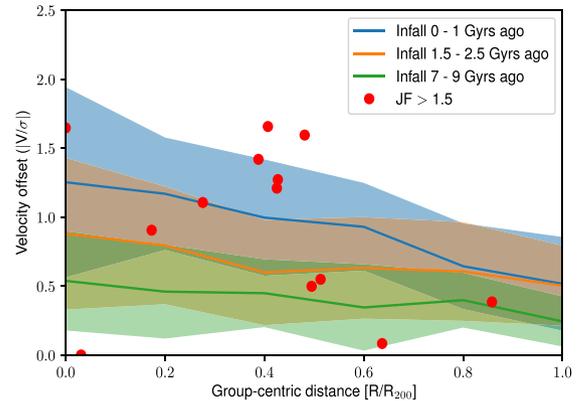

**Figure 13.** The phase-space distribution of the secure candidates, along with the expected tracks of galaxies with different infall times from the semi-analytic simulations of Henriques et al. (2020). The secure RPS candidates are also shown as red dots.

The semi-analytic galaxy formation model of Henriques et al. (2020) follows the orbits of galaxies (including satellite galaxies) using an *N*-body simulation and produces a good match to the observed stellar mass functions of galaxies over a range of redshifts. As such, its output provides a good indicator of the expected phase-space distribution of satellite galaxies in groups and clusters. Using this model, we have tracked the galaxies back through the simulation to determine how long it has been since they were accreted into the main halo. In what follows, we define the time of accretion into the main halo as when the galaxy first crosses within the three-dimensional $R_{200}$ of the halo it is in at the epoch of observation. As this time will occur between output snapshots, we use the positions and velocities of the galaxy at the snapshot before accretion to interpolate to the accretion time. We compute this accretion time for galaxies within haloes of greater than $10^{13}$ M$_\odot$ at $z = 0$.

In Fig. 13, we show the expected median and 1$\sigma$ contours for the phase-space location of galaxies with different accretion times from the simulations. We have 'observed' the simulation by projecting it along a random line of sight, so that the velocity offset and group-centric distances are in projected space, like the observations. Notice that there is a monotonic relationship between average velocity offset and infall time, such that galaxies which fell into the group within the last Gyr, they have consistently higher velocity offsets at a fixed group-centric distance than galaxies that were accreted longer ago. We also show the phase-space position of the secure RPS candidates and find that 7 of the 13 candidates fall within the 1$\sigma$ contours of the simulated galaxies that were accreted in the last Gyr. Indeed, all but two of the candidates are within the 2$\sigma$ contours. While 5 of the 13 candidates are also within the 1$\sigma$ contours of the 1.5–2.5 Gyr track, the highest velocity excess galaxies are difficult to explain with these longer infall times. This supports the conclusion that jellyfish galaxies may be on their first passage through the group, and thus are encountering the dense group medium for the first time. Note, that the point here is that they have not made multiple passages through the group, not that they have not passed the densest part of their first orbit. This would have important implications for the expected tail directions.

Our simulation results and interpretation are supported by previous results. For instance, Rhee et al. (2017) classify galaxies within $R_{200}$ and with a velocity offset greater than 1 as 'recent infallers'. By that definition, 7 of our 13 galaxies are recent infallers. The simulations of Oman et al. (2013) are also consistent with our results, although the timings are slightly different as they 'start-the-clock' on galaxy accretion when it passes $2.5R_{200}$. Notably, both of those results are based on pure *N*-body simulations, so are complimentary to our methods.

It might be wondered why not all of the RPS candidates in our sample fall in the region of recent infall as determined by the simulations. It is to be remembered that galaxy groups tend to have few members, with some of our systems having only five spectroscopically confirmed galaxies. Thus, determining a central position, a radial extent and a velocity dispersion have significant inherent uncertainties when probed by such low numbers. Thus, it is not unexpected that some of the group galaxies appear in inconsistent regions of phase space.

Given that we have found the phase-space distributions are consistent with recent infallers, we can examine the tail directions from Section 4.5 in this context. If the orbiting galaxies have yet to meet the densest component of the IGM in their path, it is likely that they will have tails pointed away from the group centre – consistent with being on their first infall. Indeed, we did see many galaxies with these characteristics. However, other RPS candidates have tails pointing towards the centre of the group. This suggests that these candidates have already passed the densest points in their orbits, and are 'on the way out' of the group. As was also seen in Roberts et al. (2021a), many of the ram-pressure-stripped candidates that are not in the infall region of a galaxy group may have passed the group pericentre, and are now backsplashing (e.g. Oman et al. 2013). Given groups have a lower ICM and velocity dispersion than clusters, the time taken to reach maximum ram-pressure stripping will be longer than in clusters (Roberts et al. 2021a), which would explain both the position of the RPS candidates in phase space, as well as the candidates with tails pointing towards the group centre. As we will see in Section 5.3, it is likely that high-mass galaxies in low-mass groups will reach low cluster radii before they are stripped of most of their gas, consistent with this picture.

### 5.3 Is ram-pressure stripping expected in groups?

Given that galaxy groups have a lower average ICM density and lower velocity dispersion than galaxy clusters, it has been previously thought that ram-pressure stripping of cold gas may not be sufficiently strong in galaxy groups efficiently strip galaxies (e.g. Roberts et al. 2021a). However, The results presented suggest that there are several galaxies within group sized haloes that have morphological deformations consistent with probable ram-pressure stripping. We can use a simple analytic calculation to find if our observation of RPS candidates in galaxy groups is consistent with the physics of cold-gas stripping via interaction with a hot, dense environment (e.g. ram-pressure stripping). The theoretical framework of ram-pressure stripping has been known since the work of Gunn & Gott (1972), and has generally been validated in hydrodynamical simulations (Abadi, Moore & Bower 1999; McCarthy et al. 2008). Thus, a robust physical model is easily created provided we have a model for the mass (and gas) distributions of the infalling galaxies and host groups.

As we have seen, the galaxies undergoing this stripping are largely star-forming and also tend to have spiral appearances. For this reason,

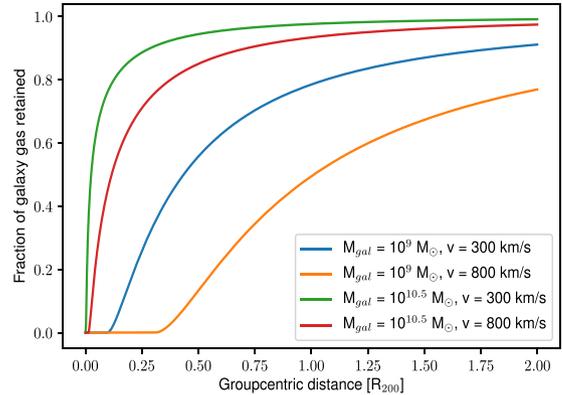

**Figure 14.** The fraction of gas retained for a galaxy with a given stellar mass and infall velocity as a function of group-centric radius using our physical model.

we assume that the disc (both the stellar and cold gas components) follow an exponential profile. For any given total stellar mass, a galaxy will have a characteristic size with relatively small scatter. In this model, we use the scaling relation for low-redshift galaxies found by Mosleh, Williams & Franx (2013) for the 'blue' galaxies. Following Jaffé et al. (2018), we assume that the gas disc has a scalelength 1.7 times the stellar disc scalelength and we also assume that the gas to stellar mass ratio is 0.5. We will discuss the effect these choices make below, but they do not qualitatively change our results. Given these choices, we can now calculate the 'restoring force' of a galaxy to itself.

We must also specify a model for the gas distribution in the host galaxy group. Galaxy groups are known to have a wide range of X-ray luminosities, and therefore a wide range of (central) gas densities. Unfortunately, the type of measurements we require are not currently available – namely, the gas density profile for optically selected groups out to at least the virial radius of the group. As such, we will assume that the gas in galaxy groups follows the dark matter profile, and that profile is described by an NFW profile (Navarro, Frenk & White 1997). Further, we will assume that the gas fraction is 0.1 $\Omega_m$.

Now that we have specified the mass and gas distributions of the galaxy and galaxy group, we can straightforwardly calculate how much gas is removed as a particular galaxy infalls due to ram pressure. Fig. 14 shows the fraction of gas that a galaxy retains as it reaches any particular group-centric distance. This is shown for two different stellar masses for the galaxy ($M_{gal} = 10^9$ and $10^{10.5}$ $M_\odot$) as well as two infalling relative velocities (300 and 800 km s$^{-1}$). The typical group velocity dispersion of the sample is about 300 km s$^{-1}$, so these are $1\sigma$ and $2.7\sigma$ velocities. It is clear from this plot that it is common for infalling galaxies in groups to have their gas removed from ram-pressure stripping. Notice that for a $10^9$-$M_\odot$ galaxy at 800 km s$^{-1}$, all of its gas has been removed by the time it reaches $0.35R_{200}$. Indeed, even for a $10^{10.5}$-$M_\odot$ galaxy at a relatively slow pace of 300 km s$^{-1}$, it will have its gas removed with $\sim 0.1R_{200}$. From this calculation, it is not at all surprising that we find ram-pressure stripping candidates in galaxy groups.

While those cases demonstrate the dependence of galaxy stripping on its infalling velocity and stellar mass, we would like to better understand the full range of those parameters. If we pick the point where half of the gas in a given galaxy has been removed (and a half has been retained) as representative of gas stripping, for each stellar mass and velocity we can calculate the group-centric radius at which that stripping occurs. In Fig. 15, we show the contours for the location of this 50 per cent stripping location (in $R_{200}$). As we can see, only the most massive and slowest moving galaxies will not see significant ram-pressure stripping in this model.

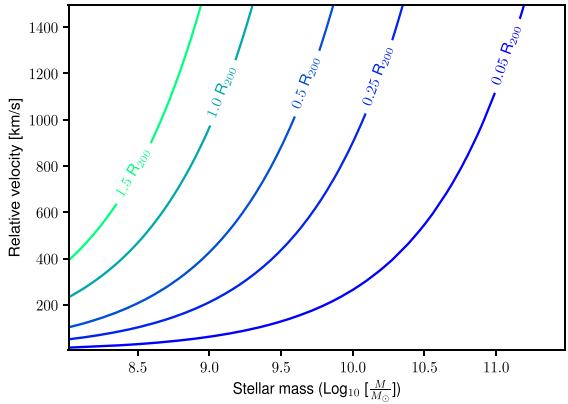

**Figure 15.** Contour plot of the groupcentric distance at which 50 per cent of the gas of a galaxy is removed (in $R_{200}$) as function of the galaxy stellar mass and relative infalling velocity within our physical model.

As the gas is stripped, it is likely that the star formation in the galaxies will quench from outside in. There are several examples of ram pressure being linked to a truncation of star formation in galaxy discs (e.g. Chung et al. 2009; Cortese et al. 2011; Vollmer et al. 2012; Gullieuszik et al. 2017).

Indeed, several spiral discs in the Virgo cluster with a truncated H $\alpha$ disc from Koopmann & Kenney (2004) have since been reported to have ram-pressure stripping features (e.g. NGC 4569, Boselli et al. 2016; NGC 4501, NGC 4522, Vollmer et al. 2012; and NGC 4299, NGC 4424, Chung et al. 2007, 2009). Analysis of the discs in our ram-pressure stripping sample may therefore be able to confirm the stripping of gas, but is beyond the scope of this work. However, we note that while many ram-pressure-stripped galaxies show signs of a truncated disc, there are several other known ram-pressure-stripped galaxies that are not classified as truncated, including NGC 4254 (Boselli et al. 2018), NGC 4654 (Vollmer et al. 2012), and NGC 4294 (Chung et al. 2007, 2009).

It is useful to check the effect of several parameter assumptions that underpin our modelling. We can use the parameters in Fig. 14 to quantify the range of differences. We have assumed that the gas to stellar mass ratio in the galaxy is 0.5, if instead, it is 1, then the location of the gas stripping will move to smaller group-centric distances. The higher gas fraction means the gas has a higher restoring force, and thus will have to experience a stronger ram pressure. The maximum average difference for the two gas fractions for any choice of the input parameters is $0.095 R_{200}$, while the smallest is $0.016 R_{200}$. We have assumed that the group gas mass fraction was 0.1, and lowering to 0.05 causes the average radial distance to move to lower values by $0.01$–$0.09 R_{200}$, while raising it to 0.15 raises the cluster-centric distance by $0.01$–$0.06 R_{200}$. Finally, we find that the relative gas to disc size has a slight effect on the location of stripping. Changing this size from 1.7 to 1.0 means slightly less gas is stripped on the outskirts of the cluster, and the point of complete stripping occurs slightly deeper within the cluster. However, the relative average locations change by only $-0.04$ to $0.01 R_{200}$. To sum up, none of the parameter choices of our model affects the conclusion that ram-pressure stripping is expected to happen within galaxy groups.

### 5.4 Prevalence of ram-pressure stripping in groups and clusters

Here we compare the number of ram-pressure-stripped galaxies found in our sample to other ram-pressure-stripped galaxies samples. Considering only our secure candidates, we have 13 from 1311 group galaxies, yielding a fraction of the total group population of $\sim$ 1 per cent. However, many other studies (e.g. Poggianti et al. 2016; Roberts & Parker 2020; Roberts et al. 2021b, see also Vulcani et al. 2022) measure the fraction of ram-pressure-stripped candidates compared to star-forming, spiral, and/or blue galaxies. We therefore measure the jellyfish fraction compared to galaxies with an sSFR $>10^{11}$ yr$^{-1}$, in line with several of these previous studies (Roberts & Parker 2020; Roberts et al. 2021a,b). While this sSFR threshold is lower than that used in Section 4.1, it will allow for the best comparison to previous studies.

We find the fraction of jellyfish galaxies compared to galaxies with $\log(\mathrm{sSFR}) > -11$ is 1.93 per cent. If we include the tentative candidates, the fraction rises to 6.69 per cent. We caution that this number is very likely an overestimate, with many marginal candidates with lower significance. This fraction is similar to that seen in Poggianti et al. (2016), who found the fraction of strong ram-pressure-stripped galaxies in optical images in WINGS clusters to be $\sim$ 2 per cent. Our fraction is however, lower than the broadband CFHT identified galaxies in the Coma cluster (4.83 per cent; Roberts & Parker 2020), and from H $\alpha$ imaging in the merging cluster Abell 901/2 (16 per cent of all H $\alpha$ emitters; Roman-Oliveira et al. 2019). In group mass haloes, Roberts et al. (2021a) find LOFAR radio tails in 1.7 per cent of group galaxies, in line with the fraction of secure candidates we have presented in this study.

### 5.5 What role in the environmental trends does ram-pressure stripping play?

We have now seen that some galaxies infalling into groups for the first time have some of their cold gas stripped directly. In this section, we discuss what implications our results might have for the general environmental transformation processes. At these redshifts, it is generally found that the quenching efficiency of satellite galaxies over central galaxies is about 40 per cent (McGee et al. 2011; Omand, Balogh & Poggianti 2014). For instance, using the GAMA quenched fractions in Davies et al. (2019), at a stellar mass of $10^{10}$ M$_\odot$ and a group mass of $10^{13}$–$10^{14}$ M$_\odot$, the fraction of quenched satellites is 0.53, while for centrals of the same stellar mass, it is 0.17. So, about 30–40 per cent of galaxies would have to be transformed by ram-pressure stripping to explain this entire difference. However, in the accretion models of McGee et al. (2009), it was shown that a group adds about 5 per cent of its galaxies each Gyr, so a continual process of at least 2–3 per cent per Gyr would be required to keep the difference the same.

Considering only our secure candidates, we have 13 from 1311 group galaxies, which is about 1 per cent of the sample. While this is clearly a much lower fraction than the total needed, a galaxy may appear ram-pressure-stripped for only a short time. The lifetimes of jellyfish phases are considerably uncertain, with observational estimates ranging from 100 Myr to 1 Gyr (Fossati et al. 2018; Bellhouse et al. 2019). This range of lifetimes would imply that 1 to 10 per cent of all galaxies would go through a jellyfish phase in a Gyr. If 10 per cent of all galaxies go through such a phase, then it would imply within only 3 or 4 Gyr 30–40 per cent of galaxies will go through such a phase. In contrast, if the lifetime was 1 Gyr, this mechanism would not be able to explain the large fraction of quenched galaxies in groups if the fraction was unchanged with redshift.

However, recall that we are only using the number of our secure candidates, which is likely a lower limit, as simple projection effects will likely hide some jellyfish features. These calculations suggest that ram-pressure stripping of this type could be the dominant effect transforming galaxies within groups. Of course, if a significant number of our ram-pressure-stripped candidates are not confirmed to be truly ram-pressure-stripped, then it limits the role of ram-pressure in galaxy groups. Further confirmation of our result is clearly needed to disentangle what is driving galaxy quenching within groups.

## 6 CONCLUSION

We have visually classified galaxies using high-quality multicolour imaging from the Hyper Suprime Cam *Subaru* Strategic Program survey according to criteria expected for galaxies undergoing active ram-pressure stripping. The examined galaxies were those within spectroscopically defined galaxy groups from the GAMA survey at redshifts $0.05 < z < 0.20$. We have also used simple physical models to understand the role of ram-pressure stripping in galaxy groups and their overall contribution to the quenching of galaxies. We summarize our findings as follows:

(i) We find 45 ram-pressure-stripped candidates from a sample of 1311 galaxies that are members of galaxy groups with halo masses between $10^{13}$ and $10^{14.6}$ M$_\odot$. Of these candidates, 13 are secure candidates with several visual features symbolic of ram-pressure stripping displayed, while the remaining 32 are tentative candidates with some ram-pressure signatures.

(ii) The RPS candidates have stellar masses across nearly the full range of the examined sample from $10^{8.5}$ and $10^{11}$ M$_\odot$. Although, at the highest stellar masses ($>10^{11}$ M$_\odot$), we do not find any RPS candidates.

(iii) The RPS candidates are predominately star forming, with 11 of the 13 secure RPS candidates having an sSFR $> 10^{-10.5}$ yr$^{-1}$. The passive galaxies are among the most massive in the sample and are near the sSFR cut-off. This may suggest that jellyfish appear if they have a given mass of gas to strip, rather than a given gas fraction. Similarly, the RPS candidates are predominately blue in the (*g* − *i*)–stellar mass plane.

(iv) In contrast to some recent studies of jellyfish in clusters, we find none of the secure RPS candidates are AGN-dominated based on their emission-line ratios. This may suggest that the ram pressure in groups is not strong enough to force gas to the central black hole.

(v) We show that the phase-space distribution of the RPS candidates is consistent with having recently been accreted into their galaxy groups ($< 1$ Gyr). Many of the ram-pressure-stripped candidates have tails pointing either away from, or towards, the group centre. This provides further evidence that these galaxies may be recent infallers, and are on radial orbits.

(vi) We also show, with the aid of a simple analytic model, that ram-pressure stripping can occur in galaxy groups. The densities towards the centre of galaxy groups are strong enough to strip gas, even with the reduced relative velocities of group galaxies. Finally, we find that given common assumptions about the time-scale for which galaxies are in the jellyfish phase, ram pressure could be the dominant form of galaxy transformation in groups.

Our results point to the potentially important role of ram-pressure stripping in environmental trends seen in galaxy groups. Future studies, with larger samples, should be able to directly elucidate the role played by galaxy stellar mass, redshift, group mass, and group state in further uncovering the role of ram pressure in groups. Future surveys using deep LSST along with eROSITA data to clarify the role of the intragroup medium will be an important development.

## ACKNOWLEDGEMENTS

TK acknowledges studentship support from Liverpool John Moores University, Faculty of Engineering and Technology and STFC. JPC and SM acknowledges support from STFC through grant number ST/N021702/1. JPC also acknowledges support from Comitee Mixto ESO-Gobierno de Chile and partial support from FONDECYT through grant 3210709.

## DATA AVAILABILITY

The data underlying this paper were accessed from the Galaxy And Mass Assembly survey (GAMA) DR3, at http://www.gama-survey.org/dr3/ and the Hyper Suprime-Cam *Subaru* Strategic Program (HSC-SSP), at https://hsc.mtk.nao.ac.jp/ssp/.

## REFERENCES

Abadi M. G., Moore B., Bower R. G., 1999, MNRAS, 308, 947
Aihara H. et al., 2018, PASJ, 70, S4
Aihara H. et al., 2019, PASJ, 71, 114
Baldry I. K. et al., 2018, MNRAS, 474, 3875
Baldwin J. A., Phillips M. M., Terlevich R., 1981, PASP, 93, 5
Balogh M. L., Morris S. L., Yee H. K. C., Carlberg R. G., Ellingson E., 1999, ApJ, 527, 54
Balogh M. L., Navarro J. F., Morris S. L., 2000, ApJ, 540, 113
Barnes J. E., Hernquist L., 1996, ApJ, 471, 115
Bellhouse C. et al., 2017, ApJ, 844, 49
Bellhouse C. et al., 2019, MNRAS, 485, 1157
Biviano A., Katgert P., Thomas T., Adami C., 2002, A&A, 387, 8
Boselli A. et al., 2016, A&A, 587, A68
Boselli A. et al., 2018, A&A, 614, A56
Butcher H., Oemler A. J., 1984, ApJ, 285, 426
Cameron E., 2011, PASA, 28, 128
Chung A., van Gorkom J. H., Kenney J. D. P., Crowl H., Vollmer B., 2009, AJ, 138, 1741
Chung A., van Gorkom J. H., Kenney J. D. P., Vollmer B., 2007, ApJ, 659, L115
Cleland C., McGee S. L., 2021, MNRAS, 500, 590
Cortese L., Catinella B., Boissier S., Boselli A., Heinis S., 2011, MNRAS, 415, 1797
Davies L. J. M. et al., 2016, MNRAS, 461, 458
Davies L. J. M. et al., 2019, MNRAS, 483, 5444
Davies R. D., Lewis B. M., 1973, MNRAS, 165, 231
Deger S. et al., 2018, ApJ, 869, 6
Dressler A., 1980, ApJ, 236, 351
Driver S. P. et al., 2011, MNRAS, 413, 971
Durret F., Chiche S., Lobo C., Jauzac M., 2021, A&A, 648, A63
Ebeling H., Stephenson L. N., Edge A. C., 2014, ApJ, 781, L40
Fasano G. et al., 2006, A&A, 445, 805
Font A. S. et al., 2008, MNRAS, 389, 1619
Fossati M. et al., 2017, ApJ, 835, 153
Fossati M. et al., 2018, A&A, 614, A57
Fossati M., Fumagalli M., Boselli A., Gavazzi G., Sun M., Wilman D. J., 2016, MNRAS, 455, 2028
Fumagalli M., Fossati M., Hau G. K. T., Gavazzi G., Bower R., Sun M., Boselli A., 2014, MNRAS, 445, 4335
Gatti M., Shankar F., Bouillot V., Menci N., Lamastra A., Hirschmann M., Fiore F., 2016, MNRAS, 456, 1073
Gavazzi G., Boselli A., Mayer L., Iglesias-Paramo J., Vílchez J. M., Carrasco L., 2001, ApJ, 563, L23

Giodini S., Lovisari L., Pointecouteau E., Ettori S., Reiprich T. H., Hoekstra H., 2013, Space Sci. Rev., 177, 247
Giovanelli R., Haynes M. P., 1985, ApJ, 292, 404
Gordon Y. A. et al., 2018, MNRAS, 475, 4223
Gullieuszik M. et al., 2017, ApJ, 846, 27
Gunawardhana M. L. P. et al., 2011, MNRAS, 415, 1647
Gunn J. E., Gott J., Richard I., 1972, ApJ, 176, 1
Gwyn S. D. J., 2012, AJ, 143, 38
Hearin A. P., Watson D. F., 2013, MNRAS, 435, 1313
Henriques B. M. B., Yates R. M., Fu J., Guo Q., Kauffmann G., Srisawat C., Thomas P. A., White S. D. M., 2020, MNRAS, 491, 5795
Heymans C. et al., 2012, MNRAS, 427, 146
Hopkins A. M. et al., 2003, ApJ, 599, 971
Jaffé Y. L. et al., 2018, MNRAS, 476, 4753
Jaffé Y. L., Smith R., Candlish G. N., Poggianti B. M., Sheen Y.-K., Verheijen M. A. W., 2015, MNRAS, 448, 1715
Joshi G. D., Pillepich A., Nelson D., Marinacci F., Springel V., Rodriguez-Gomez V., Vogelsberger M., Hernquist L., 2020, MNRAS, 496, 2673
Kauffmann G. et al., 2003, MNRAS, 346, 1055
Kawata D., Mulchaey J. S., 2008, ApJ, 672, L103
Kenney J. D. P., Geha M., Jáchym P., Crowl H. H., Dague W., Chung A., van Gorkom J., Vollmer B., 2014, ApJ, 780, 119
Kenney J. D. P., van Gorkom J. H., Vollmer B., 2004, AJ, 127, 3361
Kewley L. J., Dopita M. A., Sutherland R. S., Heisler C. A., Trevena J., 2001, ApJ, 556, 121
Koopmann R. A., Kenney J. D. P., 2004, ApJ, 613, 866
Larson R. B., Tinsley B. M., Caldwell C. N., 1980, ApJ, 237, 692
Liske J. et al., 2015, MNRAS, 452, 2087
Mahajan S., Mamon G. A., Raychaudhury S., 2011, MNRAS, 416, 2882
Man A. W. S., Toft S., Zirm A. W., Wuyts S., van der Wel A., 2012, ApJ, 744, 85
McCarthy I. G., Frenk C. S., Font A. S., Lacey C. G., Bower R. G., Mitchell N. L., Balogh M. L., Theuns T., 2008, MNRAS, 383, 593
McGee S. L., 2013, MNRAS, 436, 2708
McGee S. L., Balogh M. L., Bower R. G., Font A. S., McCarthy I. G., 2009, MNRAS, 400, 937
McGee S. L., Balogh M. L., Wilman D. J., Bower R. G., Mulchaey J. S., Parker L. C., Oemler A., 2011, MNRAS, 413, 996
McGee S. L., Bower R. G., Balogh M. L., 2014, MNRAS, 442, L105
McPartland C., Ebeling H., Roediger E., Blumenthal K., 2016, MNRAS, 455, 2994
Miyazaki S. et al., 2018, PASJ, 70, S1
Moore B., Katz N., Lake G., Dressler A., Oemler A., 1996, Nature, 379, 613
Moore B., Lake G., Katz N., 1998, ApJ, 495, 139
Mosleh M., Williams R. J., Franx M., 2013, ApJ, 777, 117
Munari E., Biviano A., Borgani S., Murante G., Fabjan D., 2013, MNRAS, 430, 2638
Muzzin A. et al., 2012, ApJ, 746, 188
Muzzin A. et al., 2014, ApJ, 796, 65
Navarro J. F., Frenk C. S., White S. D. M., 1997, ApJ, 490, 493
Noble A. G., Webb T. M. A., Muzzin A., Wilson G., Yee H. K. C., van der Burg R. F. J., 2013, ApJ, 768, 118
Oman K. A., Hudson M. J., Behroozi P. S., 2013, MNRAS, 431, 2307
Omand C. M. B., Balogh M. L., Poggianti B. M., 2014, MNRAS, 440, 843
Peluso G. et al., 2022, ApJ, 927, 130
Pierre M. et al., 2016, A&A, 592, A1
Poggianti B. M. et al., 2016, AJ, 151, 78
Poggianti B. M. et al., 2017a, Nature, 548, 304
Poggianti B. M. et al., 2017b, ApJ, 844, 48
Rasmussen J., Ponman T. J., Mulchaey J. S., 2006, MNRAS, 370, 453
Rasmussen J., Ponman T. J., Verdes-Montenegro L., Yun M. S., Borthakur S., 2008, MNRAS, 388, 1245
Rawle T. D. et al., 2014, MNRAS, 442, 196
Rhee J., Smith R., Choi H., Yi S. K., Jaffé Y., Candlish G., Sánchez-Jánssen R., 2017, ApJ, 843, 128
Ricarte A., Tremmel M., Natarajan P., Quinn T., 2020, ApJ, 895, L8
Roberts I. D. et al., 2021b, A&A, 650, A111
Roberts I. D. et al., 2022, MNRAS, 509, 1342
Roberts I. D., Parker L. C., 2020, MNRAS, 495, 554
Roberts I. D., van Weeren R. J., McGee S. L., Botteon A., Ignesti A., Rottgering H. J. A., 2021a, A&A, 650, A111
Robotham A. S. G. et al., 2011, MNRAS, 416, 2640
Roediger E., Brüggen M., 2006, MNRAS, 369, 567
Roman-Oliveira F. V., Chies-Santos A. L., Rodríguez del Pino B., Aragón-Salamanca A., Gray M. E., Bamford S. P., 2019, MNRAS, 484, 892
Sanders D. B., Soifer B. T., Elias J. H., Madore B. F., Matthews K., Neugebauer G., Scoville N. Z., 1988, ApJ, 325, 74
Sharp R. et al., 2006, in McLean I. S., Iye M., eds, Proc. SPIE Conf. Ser. Vol. 6269, Ground-based and Airborne Instrumentation for Astronomy. SPIE, Bellingham, p. 62690G
Smith R. J. et al., 2010, MNRAS, 408, 1417
Taylor E. N. et al., 2011, MNRAS, 418, 1587
Tonnesen S., Bryan G. L., 2012, MNRAS, 422, 1609
van de Voort F., Bahé Y. M., Bower R. G., Correa C. A., Crain R. A., Schaye J., Theuns T., 2017, MNRAS, 466, 3460
van der Burg R. F. J. et al., 2020, A&A, 638, A112
van Son L. A. C. et al., 2019, MNRAS, 485, 396
Vollmer B., 2003, A&A, 398, 525
Vollmer B., Wong O. I., Braine J., Chung A., Kenney J. D. P., 2012, A&A, 543, A33
Vulcani B. et al., 2018, MNRAS, 480, 3152
Vulcani B. et al., 2021, ApJ, 914, 27
Vulcani B., Poggianti B. M., Smith R., Moretti A., Jaffé Y. L., Gullieuszik M., Fritz J., Bellhouse C., 2022, ApJ, 927, 91
Weinmann S. M., van den Bosch F. C., Yang X., Mo H. J., 2006, MNRAS, 366, 2
Wetzel A. R., Tinker J. L., Conroy C., 2012, MNRAS, 424, 232
Williams B. A., Rood H. J., 1987, ApJS, 63, 265
Yoshida N., Omukai K., Hernquist L., 2008, Science, 321, 669

## APPENDIX A: TABLE OF RAM PRESSURE STRIPPING CANDIDATES

We include the GAMA Catalogue ID, Right Ascension, Declination, redshift and averaged JF class for each of our 43 RPS candidates in Table A1.

Table A1. The GAMA Catalogue ID (CATAID), Right Ascension (RA), Declination (Dec.), galaxy redshift ($z$), and averaged JF Class ($\overline{JF}$) values of 43 RPS candidates, where $\overline{JF} \geq 1.5$ and $0.5 \leq \overline{JF} < 1.5$ represent the secure and tentative candidates, respectively.

| CATAID | RA | Dec. | $z$ | $\overline{JF}$ |
|---|---|---|---|---|
| 1775685 | 36.29 | −5.10 | 0.09 | 3.0 |
| 1446793 | 31.50 | −4.04 | 0.14 | 3.0 |
| 1213244 | 34.01 | −4.53 | 0.13 | 2.3 |
| 2408024 | 37.86 | −5.35 | 0.14 | 2.3 |
| 2234271 | 31.22 | −4.63 | 0.11 | 2.0 |
| 1319522 | 32.83 | −4.91 | 0.14 | 2.0 |
| 2335347 | 36.21 | −5.68 | 0.05 | 2.0 |
| 2333318 | 36.45 | −6.02 | 0.05 | 2.0 |
| 1214289 | 33.69 | −4.48 | 0.14 | 1.6 |
| 2338305 | 36.23 | −5.14 | 0.08 | 1.6 |
| 2334142 | 36.25 | −5.88 | 0.05 | 1.6 |
| 1436898 | 31.19 | −4.68 | 0.1 | 1.6 |
| 1775219 | 36.82 | −5.13 | 0.14 | 1.6 |
| 1322096 | 32.96 | −4.73 | 0.07 | 1.3 |
| 1900600 | 37.67 | −4.39 | 0.14 | 1.3 |
| 2230644 | 31.71 | −5.35 | 0.14 | 1.3 |
| 1668316 | 35.31 | −4.92 | 0.14 | 1.3 |
| 2165364 | 33.67 | −4.58 | 0.14 | 1.0 |
| 2234028 | 31.18 | −4.68 | 0.11 | 1.0 |
| 2305411 | 35.29 | −4.65 | 0.08 | 1.0 |
| 2304513 | 35.38 | −4.81 | 0.15 | 1.0 |
| 2305404 | 35.18 | −4.64 | 0.2 | 1.0 |
| 2005204 | 38.59 | −4.92 | 0.14 | 1.0 |
| 2305602 | 35.34 | −4.62 | 0.08 | 1.0 |
| 2344381 | 36.34 | −4.07 | 0.17 | 1.0 |
| 2131428 | 34.06 | −4.24 | 0.15 | 1.0 |
| 1101008 | 34.16 | −4.33 | 0.15 | 1.0 |
| 2000072 | 37.89 | −5.23 | 0.14 | 1.0 |
| 1673619 | 35.39 | −4.57 | 0.16 | 1.0 |
| 1321282 | 32.96 | −4.78 | 0.07 | 1.0 |
| 1326493 | 32.42 | −4.42 | 0.14 | 1.0 |
| 1418229 | 31.81 | −5.91 | 0.09 | 1.0 |
| 1418630 | 31.81 | −5.89 | 0.09 | 1.0 |
| 1426399 | 31.64 | −5.39 | 0.14 | 1.0 |
| 1215214 | 33.73 | −4.40 | 0.14 | 1.0 |
| 1545810 | 30.41 | −5.54 | 0.19 | 1.0 |
| 1213397 | 33.62 | −4.53 | 0.14 | 1.0 |
| 2266760 | 30.78 | −5.30 | 0.13 | 0.6 |
| 1213740 | 33.63 | −4.50 | 0.14 | 0.6 |
| 1675008 | 35.25 | −4.48 | 0.2 | 0.6 |
| 1761212 | 36.29 | −6.07 | 0.05 | 0.6 |
| 2196318 | 32.42 | −5.33 | 0.14 | 0.6 |
| 2165561 | 33.73 | −4.56 | 0.14 | 0.6 |
| 1539734 | 30.60 | −5.91 | 0.19 | 0.5 |
| 1998493 | 37.90 | −5.33 | 0.14 | 0.5 |

This paper has been typeset from a T<sub>E</sub>X/L<sup>A</sup>T<sub>E</sub>X file prepared by the author.